\makeatletter
\input{filecontents.sty}
\makeatother

\documentclass[aip,ajp,reprint,secnumarabic,groupedaddress]{preprint}

\renewcommand\documentclass[2][]{}
\documentclass[aip,jap,reprint,groupedaddress]{revtex4-1}

\usepackage{graphicx}
\usepackage{color}
\usepackage{grffile}
\usepackage{amsmath}
\usepackage{amssymb}
\usepackage{amsthm}
\usepackage{braket}
\usepackage{textcomp}
\usepackage{microtype}
\usepackage[
breaklinks=true,
pdfauthor={Heiko Bauke and Noya Ruth Itzak},
pdftitle={Visualizing quantum mechanics in phase space}
]{hyperref}
\usepackage{txfonts}
\usepackage{comment}

\graphicspath{{figures/}}

\renewcommand{\vec}[1]{\boldsymbol{#1}}
\newcommand{\D}{\mathrm{d}}
\newcommand{\I}{\mathrm{i}}
\newcommand{\e}{\mathrm{e}}
\begin{document}

\title{Visualizing quantum mechanics in phase space}
\author{Heiko Bauke} 
\email{bauke@mpi-hd.mpg.de}
\author{Noya Ruth Itzhak} 
\affiliation{Max-Planck-Institut f{\"u}r Kernphysik, Saupfercheckweg~1,
  69117~Heidelberg, Germany}

\date{\today}

\begin{abstract}
  We examine the visualization of quantum mechanics in phase space by
  means of the Wigner function and the Wigner function flow as a
  complementary approach to illustrating quantum mechanics in
  configuration space by wave functions. The Wigner function formalism
  resembles the mathematical language of classical mechanics of
  non-interacting particles. Thus, it allows a more direct comparison
  between classical and quantum dynamical features.
\end{abstract}


\maketitle

\section{Introduction}
\label{sec:intro}

Quantum mechanics is a corner stone of modern physics and technology.
Virtually all processes that take place on an atomar scale require a
quantum mechanical description. For example, it is not possible to
understand chemical reactions or the characteristics of solid states
without a sound knowledge of quantum mechanics. Quantum mechanical
effects are utilized in many technical devices ranging from transistors
and Flash memory to tunneling microscopes and quantum cryptography, just
to mention a few applications.

Quantum effects, however, are not directly accessible to human senses,
the mathematical formulations of quantum mechanics are abstract and its
implications are often unintuitive in terms of classical physics. This
poses a major challenge in teaching and learning the foundations of
quantum mechanics and makes it difficult for students to form a mental
picture of quantum dynamical processes.  Therefore, teaching quantum
mechanics may be supported by visualizing the mathematical objects of
quantum mechanics' mathematical language in order to sharpen the
students' understanding of abstract concepts. In fact, many text books
\cite{Thaller:2000:vis_QM,Thaller:2005:vis_QM_II,Robinett:2006:QM,Brandt:Dahmen:2001:Picture_book}
communicate quantum mechanics with the aid of visual means.

Visualizing quantum dynamics is typically implemented by providing a
pictorial representation of the quantum mechanical wave function in
position space \cite{Goldberg:etal:1967:Motion_Pictures} or in momentum
space \cite{Goldberg:etal:1968:Scattering}. In order to establish a
relation between quantum dynamics and classical dynamics, the phase
space seems to be more appropriate because classical Hamiltonian
mechanics is formulated in phase space. Thus, we will present in this
contribution methods to visualize quantum dynamics in phase space.

The remainder of this paper is organized as follows. In
section~\ref{sec:pos_mom_space} we briefly review how to depict quantum
dynamics in position space or in momentum space using wave functions. In
section~\ref{sec:qps} we give a short introduction to quantum mechanics
in phase space and establish some connections between quantum mechanics
and many-particle classical mechanics before we visualize quantum
dynamics in phase space in section~\ref{sec:vqps}.

\section{Visualizing quantum dynamics in position space or momentum space} 
\label{sec:pos_mom_space}

\begin{figure}
  \centering
  \includegraphics[scale=0.45]{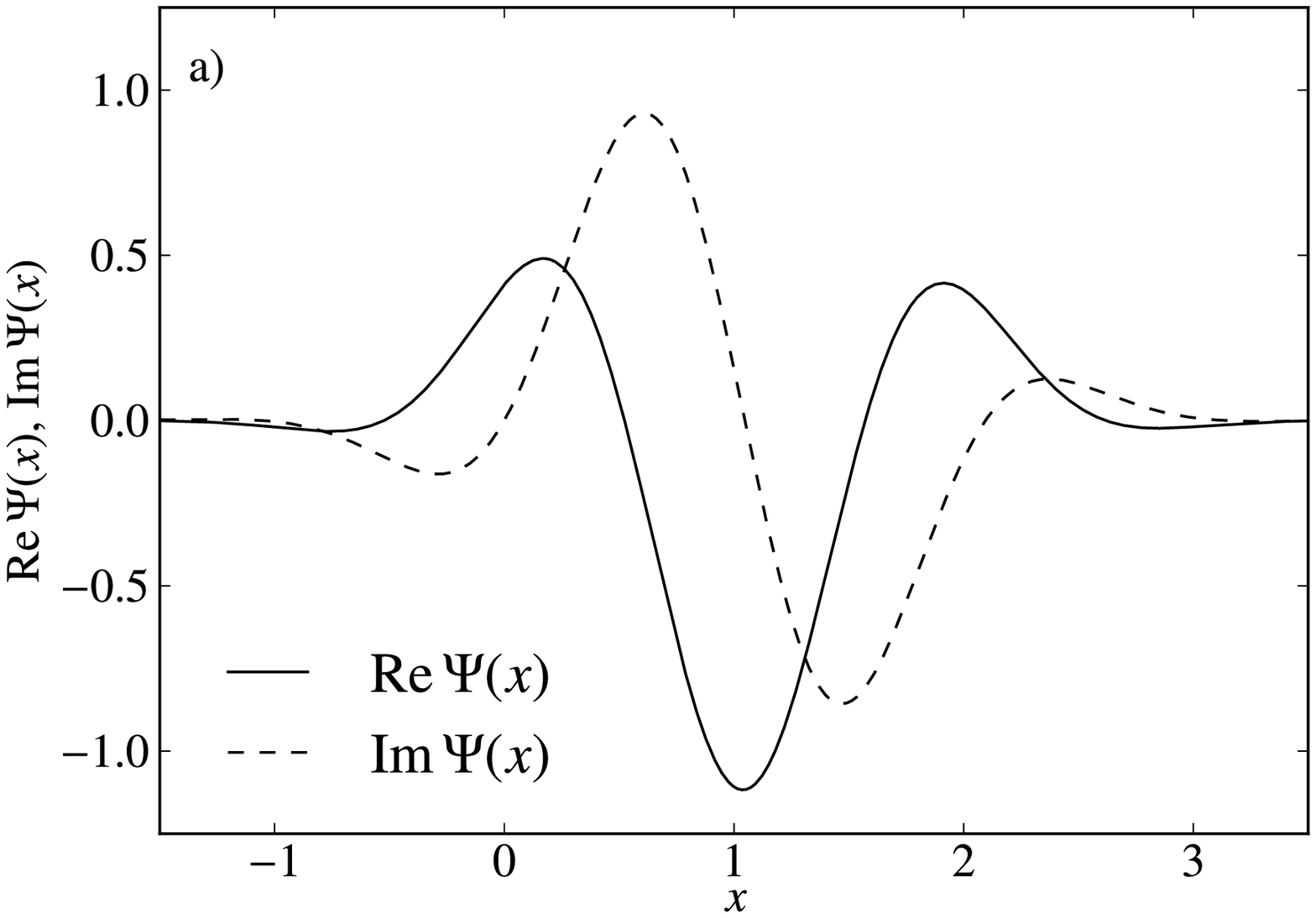}\\[1ex]%
  \includegraphics[scale=0.45]{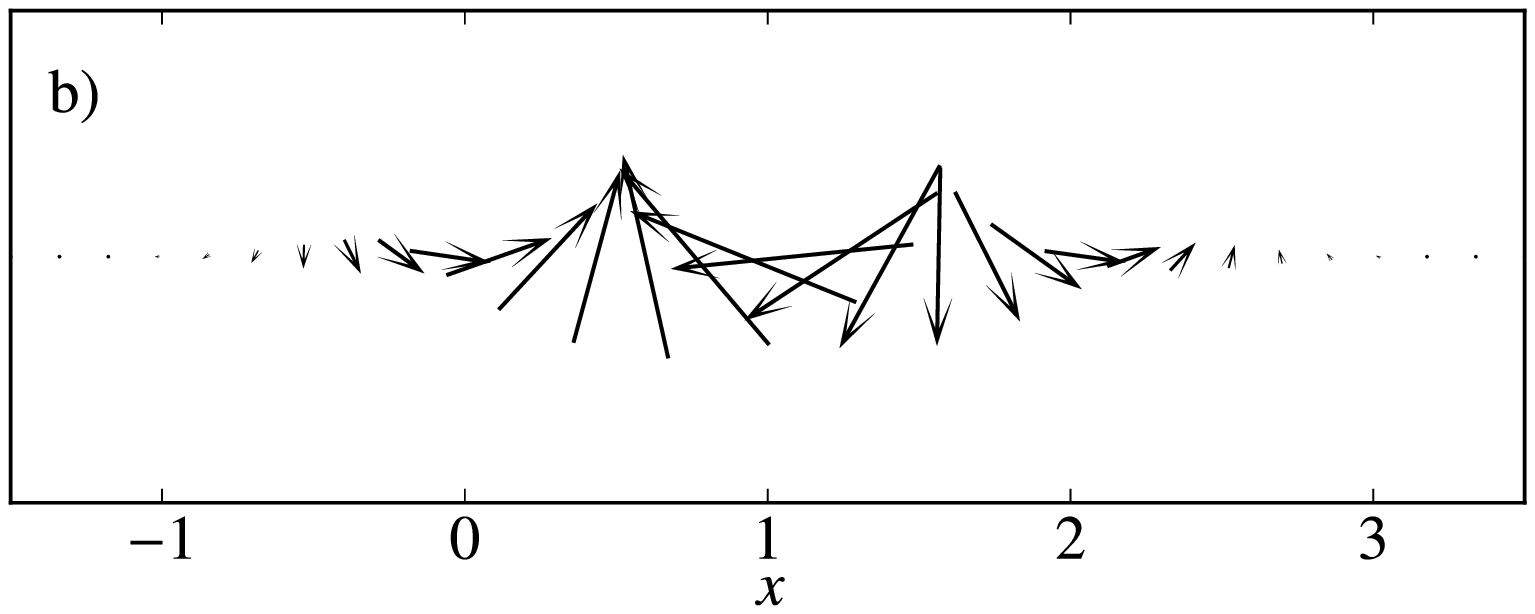}\\[0.2ex]%
  \includegraphics[scale=0.45]{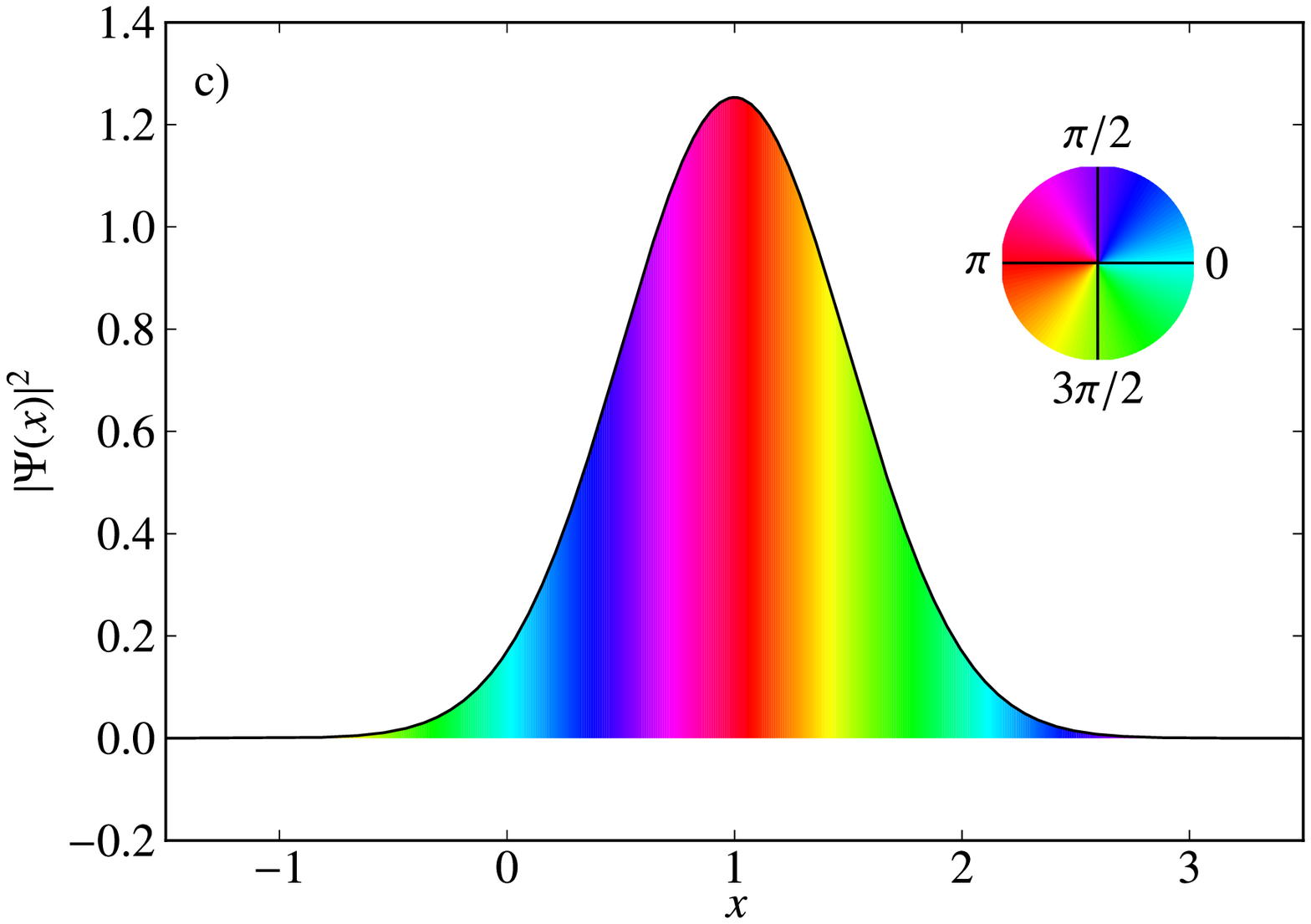}%
  \caption{(Color online) Three different ways of visualizing the
    Gaussian wave packet (\ref{eq:gauss}) at time $t=0$ with $\bar x=1$,
    $\bar p=3$, and $\sigma=1/2$ (dimensionless units with $\hbar=1$ and
    $m=1$); a) visualization by the wave-function's real part and
    imaginary part, b) visualization by the wave-function's modulus and
    its phase using arrows (phasors), c) visualization by plotting the
    wave-function's squared modulus and encoding the phase in a periodic
    color scheme.}
  \label{fig:free_particle_vis}
\end{figure}

A (one-dimensional) quantum mechanical position space wave function maps
each point $x$ on the position axis to a time dependent complex number
$\Psi(x, t)$.  Equivalently, one may consider the wave function
$\tilde\Psi(p, t)$ in momentum space, which is given by a Fourier
transform of $\Psi(x, t)$, viz.
\begin{equation}
  \label{eq:Psi_momentum}
  \tilde\Psi(p, t) = 
  \frac{1}{(2\pi\hbar)^{1/2}} \int \Psi(x, t) \e^{-\I x p/\hbar}\D x\,.
\end{equation}
Various methods to visualize complex-valued functions in position space
or momentum space have been devised.  Figure~\ref{fig:free_particle_vis}
exemplifies three different ways of visualizing a wave function. It
shows a Gaussian wave packet
\begin{multline}
  \label{eq:gauss}
  \Psi(x, t) = 
  \frac{1}{\left(
      2\pi\sigma^2\left(
        1+\frac{\I\hbar t}{2m\sigma^2}
      \right)^2
    \right)^{1/4}}\\
  \exp\left(
    \left( -
      \frac{(x-\bar x)^2}{4\sigma^2 } + 
      \frac{\I \bar p (x-\bar x)}{\hbar} -
      \frac{\I \bar p^2t}{2m\hbar} 
    \right)\Bigg/\left(1+\frac{\I\hbar t}{2m\sigma^2}\right)
  \right)
\end{multline}
in position space with mass $m$, mean position $\bar x$, mean momentum
$\bar p$, and width $\sigma$ at time $t=0$.

Part~a of Fig.~\ref{fig:free_particle_vis} presents the complex wave
function (\ref{eq:gauss}) by plotting its real and its imaginary part as
separate real-valued functions. Although a complex function is
completely determined by specifying its real and its imaginary part it
is difficult to infer physically relevant information from these
separated quantities.

Visualizing the wave-function's squared modulus and its phase is more
appropriate because the squared modulus of a wave function may be
interpreted as a probability distribution and the change of its phase
encodes information about the momentum. Thus, we picture the wave
function (\ref{eq:gauss}) in part~b of Fig.~\ref{fig:free_particle_vis}
by small directional arrows. Each arrow's length is determined by the
wave-function's modulus, its direction (angle between the arrow and the
horizontal axis) by the wave-function's phase. The arrows in
Fig.~\ref{fig:free_particle_vis}~b are sometimes called phasors. The
method to portray a wave function by phasors had been used, for example,
by Feynman \cite{Feynman:1986:QED}. It is beneficial in situations where
advanced printing methods, color printing for example, are not
available. However, it is difficult to picture fast oscillatory wave
functions in this manner.\enlargethispage{0.75ex}%

Another way of presenting a wave function by its modulus and its phase
has been popularized by Thaller
\cite{Thaller:2000:vis_QM,Thaller:2005:vis_QM_II}. This method is
utilized in part~c of Fig.~\ref{fig:free_particle_vis}. It shows the
squared modulus of the wave function (\ref{eq:gauss}) plus its phase
which is encoded by the color between the graph of the squared modulus
and the horizontal axis. Plotting the squared modulus allows for a
direct read off of the probability distribution in position space. Note
that the color scheme for the phase in part~c of
Fig.~\ref{fig:free_particle_vis} is \mbox{$2\pi$-periodic} and
contiguous.  Encoding the phase as a color allows to plot even
relatively strongly oscillating wave functions. It requires, however,
high-quality color printing for good reproduction.  Thaller
\cite{Thaller:2000:vis_QM,Thaller:2005:vis_QM_II} has also generalized
this approach to multi-component wave functions.

\section{Quantum mechanics in phase space}
\label{sec:qps} 

\subsection{The Wigner function}

The Wigner distrubution function
\cite{Wigner:1932:Quantum_Correction,Hillery:etal:1984:Distribution_functions,Zachos:etal:Quantum_Mechanics_in_Phase_Space}
was introduced by Eugene Paul Wigner to take into account quantum
corrections to classical statistical mechanics in 1932. Today, it finds
applications in current research in various branches of physics ranging
from quantum optics \cite{Schleich:2001:Quantum_optics} and quantum
chaos \cite{Habib:etal:1998:Decoherence} to nuclear and solid state
physics.  The Wigner function allows for a mathematical description of
(non-relativistic) quantum mechanics in phase space. But it is more than
a sheer mathematical object, it may be reconstructed by a series of
measurements
\cite{Smithey:etal:1993:Measurement_Wigner_distribution,Kurtsiefer:etal:1997:Measurement_Wigner,Breitenbach:etal:1997:Measurement_quantum_states,Lvovsky:etal:2001:Quantum_State_Reconstruction,Deleglise:etal:2008:Reconstruction}.
The Wigner function formalism is mathematically equivalent to the
standard language of quantum mechanics in terms of wave functions and
the Schr{\"o}dinger equation but it is more closely related to classical
mechanics because it formulates quantum mechanics in phase space as
Hamiltonian mechanics does.

The Wigner function of a (one-dimensional) pure state wave function
$\Psi(x, t)$ is defined as
\begin{equation}
  \label{eq:wigner_def}
  w(x, p, t) = 
  \frac{1}{2\pi\hbar} \int \Psi^*(x+s/2, t)\Psi(x-s/2, t)
  \e^{\I s p/\hbar}
  \D s\,,
\end{equation}
where $\Psi^*(x, t)$ denotes the complex conjugate of $\Psi(x, t)$.
Note that due to the symmetry of the integrand in (\ref{eq:wigner_def}),
the Wigner function $w(x, p, t)$ is always a real function.  The
expression 
\begin{equation}
  \label{eq:wigner_def2}
  w(x, p, t) = 
  \frac{1}{2\pi\hbar} \int \tilde\Psi^*(p+s/2, t)\tilde\Psi(p-s/2, t)
  \e^{-\I s x/\hbar}
  \D s
\end{equation}
is equivalent to (\ref{eq:wigner_def}) if $\Psi(x, t)$ and
$\tilde\Psi(p, t)$ are linked via (\ref{eq:Psi_momentum}). Further
important mathematical properties
\cite{Hillery:etal:1984:Distribution_functions,Schleich:2001:Quantum_optics}
of the Wigner function include the following relations.
\begin{itemize}
\item If $\Psi(x, t)$ is normalized to one then also $w( x, p, t)$, viz.
  \begin{equation}
    \iint w( x,  p, t)\,\D p\,\D x = 1\,,
  \end{equation}
  furthermore, the equation 
  \begin{equation}
    \label{eq:w2}
    \iint w( x,  p, t)^2\,\D p\,\D x = \frac{1}{2\pi\hbar}
  \end{equation}
  holds for all pure states $\Psi(x, t)$.
\item Quantum mechanical probability densities in position space and
  momentum space may be obtained from the marginals of the Wigner
  function
  \begin{subequations}
    \label{ew:w_marginal}
    \begin{align}
      \label{ew:w_marginal_x}
      \rho(x, t) = |\Psi(x, t)|^2 & = \int w(x, p, t)\,\D p \,,\\
      \label{ew:w_marginal_p}
      \rho(p, t) = |\tilde\Psi(p, t)|^2 & = \int w(x, p, t)\,\D x\,.
    \end{align}
  \end{subequations}
\item The state overlap of two wave functions $\Psi_1(x, t)$ and
  $\Psi_2(x, t)$ in terms of their Wigner functions $w_1(x, p, t)$ and
  $w_2(x, p, t)$ reads
  \begin{multline}
    \int \Psi_1^*(x, t)\Psi_2(x, t) \,\D x = \\
    2\pi\hbar\iint w_1(x, p, t)w_2(x, p, t) \,\D p\,\D x\,.
  \end{multline}
\item The Wigner function obeys the reflection symmetries
  \begin{align}
    \Psi( x, t) \to \Psi^*( x, t) & \quad\Rightarrow\quad
    w( x,  p, t) \to w( x, - p, t)\,, \\
    \Psi( x, t) \to \Psi(- x, t) & \quad\Rightarrow\quad
    w( x,  p, t) \to w(- x, - p, t)\,.
  \end{align}
\item The Wigner function is Galilei-covariant, that is
  \begin{multline}
    \Psi(x, p, t) \to \Psi(x+y, p, t)
    \quad\Rightarrow\\
    w( x,  p, t) \to w( x+ y,  p, t) \,,
  \end{multline}
  but not Lorentz-covariant. However, Lorentz-covariant generalizations
  of the Wigner function have been considered in the literature
  \cite{Bialynicki-Birula:etal:1991:Dirac,Shin:etal:1992:Wigner_function_1/2,Varro:Javanainen:2003:relativistic_Wigner_functions}
  which may be applied in relativistic quantum optics
  \cite{Keitel:2001:Relativistic_quantum_optics}.
\item The Wigner function is not gauge invariant. The definition
  (\ref{eq:wigner_def}), however, may be generalized
  \cite{Serimaa:etal:1986:Gauge-independent} such that it becomes
  independent of the gauge in the presence of a magnetic field.
\end{itemize}

The Wigner function $w(x, p, t)$ is called \emph{quasi} probability
function because it is real and normalized to one but not strictly
non-negative. In fact, a sufficient and necessary condition that $w( x,
p, t)$ is strictly non-negative is that $\Psi( x, t)$ is a minimum
uncertainty wave packet, that is a Gaussian wave packet
\cite{Hudson:1974:Wigner_non-negative}. As a consequence of
(\ref{eq:w2}), the Wigner function can not be arbitrarily highly peaked,
which is a reminiscence of the quantum mechanical uncertainty relation.

The evolution equation
\cite{Moyal:1949:Quantum_mechanics,Hillery:etal:1984:Distribution_functions,Peres:1993:Quantum_Theory,Schleich:2001:Quantum_optics}
of the Wigner function may be obtained by calculating the
Wigner-function's time derivative.  Taking into account the Schr{\"o}dinger
equation
\begin{equation}
  \label{eq:Schrodinger}
  \I\hbar\frac{\partial\Psi(x, t)}{\partial t} = \hat H\Psi(x, t) =
  -\frac{\hbar^2}{2m}\frac{\partial^2\Psi(x, t)}{\partial x^2} + 
  V(x, t)\Psi(x, t)
\end{equation}
for the wave function moving in the potential $V(x, t)$ we get
\begin{multline}
  \label{eq:dt_wig}
  \I\hbar\frac{\partial w(x, p, t)}{\partial t} = \\
  \frac{1}{2\pi\hbar}\int\left(
    \big(\hat H\Psi(x, t)\big)\Psi(x, t)^* - 
    \Psi(x, t)\big(\hat H\Psi(x, t)\big)^*
  \right)\e^{\I sp/\hbar}\,\D s \,.
\end{multline}
After several algebraic manipulations that include expanding the
potential $V(x, t)$ into a Taylor series around \mbox{$x=0$} and
integrating (\ref{eq:dt_wig}) by parts one finally finds the so-called
quantum Liouville equation
\begin{multline}
  \label{eq:qLiouville}
  \frac{\partial w(x, p, t)}{\partial t} = \\
  \sum_{l=0}^\infty
  \frac{(\I\hbar/2)^{2l}}{(2l+1)!}
  \left(
    \frac{\partial}{\partial p_{w}}
    \frac{\partial}{\partial x_{H}} -
    \frac{\partial}{\partial p_{H}}
    \frac{\partial}{\partial x_{w}}
  \right)^{2l+1} H(x, p, t) w(x, p, t) \,,
\end{multline}
which is a partial differential equation of infinite degree. The
operator $\left( \frac{\partial}{\partial x_{w}}
  \frac{\partial}{\partial p_{H}} - \frac{\partial}{\partial p_{H}}
  \frac{\partial}{\partial x_{w}} \right)$ in (\ref{eq:qLiouville})
consists of differential operators that act on the Hamilton function
$H(x, p, t)$ (indicated by the index $H$) and differential operators
that act on the Wigner function $w(x, p, t)$ (indicated by the index
$w$). The function $H(x, p, t)$ in (\ref{eq:qLiouville}) is the
classical Hamilton function 
\begin{equation}
  H(x, p, t) = \frac{p^2}{2m} + V(x, t)
\end{equation}
that corresponds to the Hamilton operator
$\hat H$ of the Schr{\"o}dinger equation (\ref{eq:Schrodinger}).

For systems where magnetic fields are absent as in
(\ref{eq:Schrodinger}), the Hamilton function $H(x, p, t)$ can be
written as a sum of a kinetic energy term $K(p)=p^2/(2m)$ that depends
only on the momentum $p$ and a potential energy term $V(x, t)$ that
depends only on the position $x$ (and possibly the time $t$). In this
case, the quantum Liouville equation (\ref{eq:qLiouville}) may be written
as a continuity equation
\begin{equation}
  \label{eq:continuity_equation}
  \frac{\partial w(x, p, t)}{\partial t} = 
  -
  \begin{pmatrix}
    \frac{\partial}{\partial x} \\[1.5ex]
    \frac{\partial}{\partial p} 
  \end{pmatrix}
  \cdot\vec j(x, p, t) \,,
\end{equation}
where we have introduced the two-dimensional vector field
\begin{equation}
  \label{eq:Wigner_flow}
  \vec j(x, p, t) =
  \begin{pmatrix}
    \frac{p}{m}w(x, p, t) \\[1.25ex]
    -\sum\limits_{l=0}^\infty
    \frac{(\I\hbar/2)^{2l}}{(2l+1)!}
    \left(
      \frac{\partial}{\partial p}
      \frac{\partial}{\partial x_{V}} -
      \frac{\partial}{\partial x_{w}}
    \right)^{2l}       
    \frac{\partial V(x, t)}{\partial x} w(x, p, t)
  \end{pmatrix}\,,
\end{equation}
called the Wigner function flow. 

Let $\hat O_x$ denote some operator function of the quantum mechanical
position operator and $O(x)$ the corresponding classical function, then
it follows from (\ref{ew:w_marginal_p}) that
\begin{equation}
  \braket{O_x}(t) =  
  \Braket{\Psi(x, t) | \hat O_x | \Psi(x, t) } = 
  \iint w(x, p, t)O(x)\,\D p \,\D x\,.
\end{equation}
This means one can calculate the quantum mechanical expectation value
$\braket{O_x}(t)$ of an operator $\hat O_x$ by averaging the classical
function $O(x)$ over the Wigner function $w(x, p, t)$. For momentum
dependent operators $\hat O_p$ and their corresponding classical
functions one finds using (\ref{ew:w_marginal_x}) a similar statement
\begin{equation}
  \braket{O_p}(t) =  
  \Braket{\Psi(x, t) | \hat O_p | \Psi(x, t) } = 
  \iint w(x, p, t)O(p)\,\D p \,\D x\,.
\end{equation}
For general operators $\hat O_{x, p}$ and classical functions $O(x, p)$
that depend on the position as well as on the momentum one can show
\cite{Moyal:1949:Quantum_mechanics,Hillery:etal:1984:Distribution_functions}
that
\begin{multline}
  \label{eq:observable}
  \braket{O_{x,p}}(t) =  
  \Braket{\Psi(x, t) | \hat O_{x, p} | \Psi(x, t) } = \\
  \iint w(x, p, t)O(x, p)\,\D p \,\D x
\end{multline}
provided that the relation between the operator $\hat O_{x, p}$ and the
function $O(x, p)$ is established by
\begin{equation}
  \hat O_{x, p} = 
  \iint \tilde O(\xi, \pi) \e^{\I(\xi\hat x + \pi\hat p)/\hbar}
  \,\D\xi\,\D\pi \,,
\end{equation}
where $\tilde O(\xi, \pi)$ denotes the Fourier transform of $O(x, p)$
\begin{equation}
  \tilde O(\xi, \pi) = 
  \iint O(x, p) \e^{-\I(x\xi + p\pi)/\hbar}\,\D x\,\D p
\end{equation}
and $\hat x$ the position operator and $\hat p$ the canonical momentum
operator, respectively.

The Wigner function $w_E(x, p)$ of an energy eigen-state wave function of
the Hamiltonian $H(x, p)=p^2/(2m)+V(x)$ is time independent and
therefore it follows from (\ref{eq:qLiouville}) that
\begin{subequations}
  \label{eq:Wigner_eigen}
  \begin{equation}
    \label{eq:Wigner_eigen1}
    \left(
      \frac{p}{m}\frac{\partial }{\partial x} -
      \sum_{l=0}^\infty
      \frac{(\I\hbar/2)^{2l}}{(2l+1)!}
      \frac{\partial^{2l+1} V(x)}{\partial x^{2l+1}} 
      \frac{\partial^{2l+1}}{\partial p^{2l+1}} 
    \right)
    w_E(x, p)=0 \,.
  \end{equation}
  This equation, however, is not sufficient to specify the energy
  eigen-value $E$, which is determined by the eigen-value equation
  \cite{Schleich:2001:Quantum_optics}
  \begin{multline}
    \label{eq:Wigner_eigen2}
    \left( \frac{p^2}{2m}- \frac{\hbar^2}{8m}\frac{\partial^2 }{\partial
        x^2} + \sum_{l=0}^\infty \frac{(\I\hbar/2)^{2l}}{(2l)!}
      \frac{\partial^{2l} V(x)}{\partial x^{2l}}
      \frac{\partial^{2}}{\partial p^{2l}} \right) w_E(x, p) = \\
    Ew_E(x, p) \,.
  \end{multline}
\end{subequations}
The two equations (\ref{eq:Wigner_eigen1}) and (\ref{eq:Wigner_eigen2})
together are equivalent to the time independent Schr{\"o}dinger eigen-value
equation
\begin{equation}
  -\frac{\hbar^2}{2m}\frac{\partial^2\Psi_E(x, t)}{\partial x^2} + 
  V(x, t)\Psi_E(x, t) = E \Psi_E(x, t)\,.
\end{equation}

\subsection{Relations to classical mechanics}
\label{sec:cps} 

The motion of an ensemble of classical non-interacting particles may be
characterized by a probability distribution $w(x, p, t)$ in phase space.
\footnote{We do not distinguish in our notation between the Wigner
  function and the classical probability distribution. It should be
  clear from the context whenever $w(x, p, t)$ denotes a Wigner function
  or a classical probability distribution.} The probability distribution
$w(x, p, t)$ determines how likely it is that a particle in the ensemble
is at position $x$ and has momentum $p$ at time $t$. It allows to
calculate expectation values of observables for the classical ensemble
by an integral over the phase space
\begin{equation}
  \braket{O_{x,p}}(t) = \iint w(x, p, t)O(x, p)\,\D p \,\D x
\end{equation}
likewise for a quantum system by (\ref{eq:observable}).  The evolution
of $w(x, p, t)$ is determined by the (classical) Liouville equation
\cite{Gaspard:1998:Chaos,Schwabl:2006:stat_mech}
\begin{equation}
  \label{eq:Liouville}
  \frac{\partial w(x, p, t)}{\partial t} =
  \frac{\partial H(x, p, t)}{\partial p}
  \frac{\partial w(x, p, t)}{\partial x} -
  \frac{\partial H(x, p, t)}{\partial x}
  \frac{\partial w(x, p, t)}{\partial p} \,,
\end{equation}
where $H(x, p, t)$ denotes the Hamilton function that also governs the
motion of the ensemble's single particles.  The equation
(\ref{eq:Liouville}) may be written as a continuity equation
(\ref{eq:continuity_equation}) by introducing the probability flow
\begin{equation}
  \label{eq:class_flow}
  \vec j(x, p, t) =
  \begin{pmatrix}
    \frac{p}{m}w(x, p, t) \\[1.25ex]
    -\frac{\partial V(x, t)}{\partial x} w(x, p, t)
  \end{pmatrix}\,.
\end{equation}

The quantum Liouville equation (\ref{eq:qLiouville}) reduces in the
limit $\hbar\to 0$ to the classical Liouville equation
(\ref{eq:Liouville}).  Note, however, that also if the third spatial
derivative of the potential $V(x, t)$ and all its higher derivatives
vanish the quantum Liouville equation (\ref{eq:qLiouville}) also reduces
to the classical Liouville equation (\ref{eq:Liouville}). This means
that the quantum mechanical evolution of the Wigner function for a
particle in a homogeneous constant potential or in a harmonic oscillator
(or a combination of both) follows the same dynamical law as the
probability density of an ensemble of classical particles in the same
potential. Nevertheless, non-classical features may be present in the
Wigner function.

In summary, quantum theory in phase space in terms of the Wigner
function features formal parallels to the classical theory for an
ensemble of classical non-interacting particles. The state of both kinds
of systems is completely characterized by a real-valued phase space
distribution function $w(x, p, t)$.  The evolution of $w(x, p, t)$ is
determined by the classical Liouville equation (\ref{eq:Liouville}) or
the quantum Liouville equation (\ref{eq:qLiouville}), respectively. The
quantum Liouville equation equals its classical counterpart plus some
additional quantum terms.  For classical systems, $w(x, p, t)$ may be
any non-negative normalizable function. The quantum mechanical Wigner
function, however, has to be compatible with the laws of quantum
mechanics, e.\,g., to respect the uncertainty relation, and may be
negative.  Thus, proper quantum characteristics of a quantum dynamical
process are governed the additional terms in the quantum Liouville
equation or the negative parts of the Wigner function. In fact, the
negative parts of the Wigner function are a common measure for
non-classicality
\cite{Kenfack:Zyczkowski:2004:Negative_Wigner,Kurtsiefer:etal:1997:Measurement_Wigner,Lvovsky:etal:2001:Quantum_State_Reconstruction,Mahmoudi:etal:2005:quantum_signatures}.

\begin{figure*}
  \centering
  \includegraphics[scale=0.45]{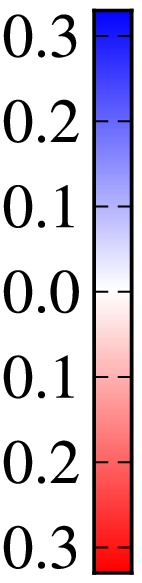}\quad
  \includegraphics[scale=0.45]{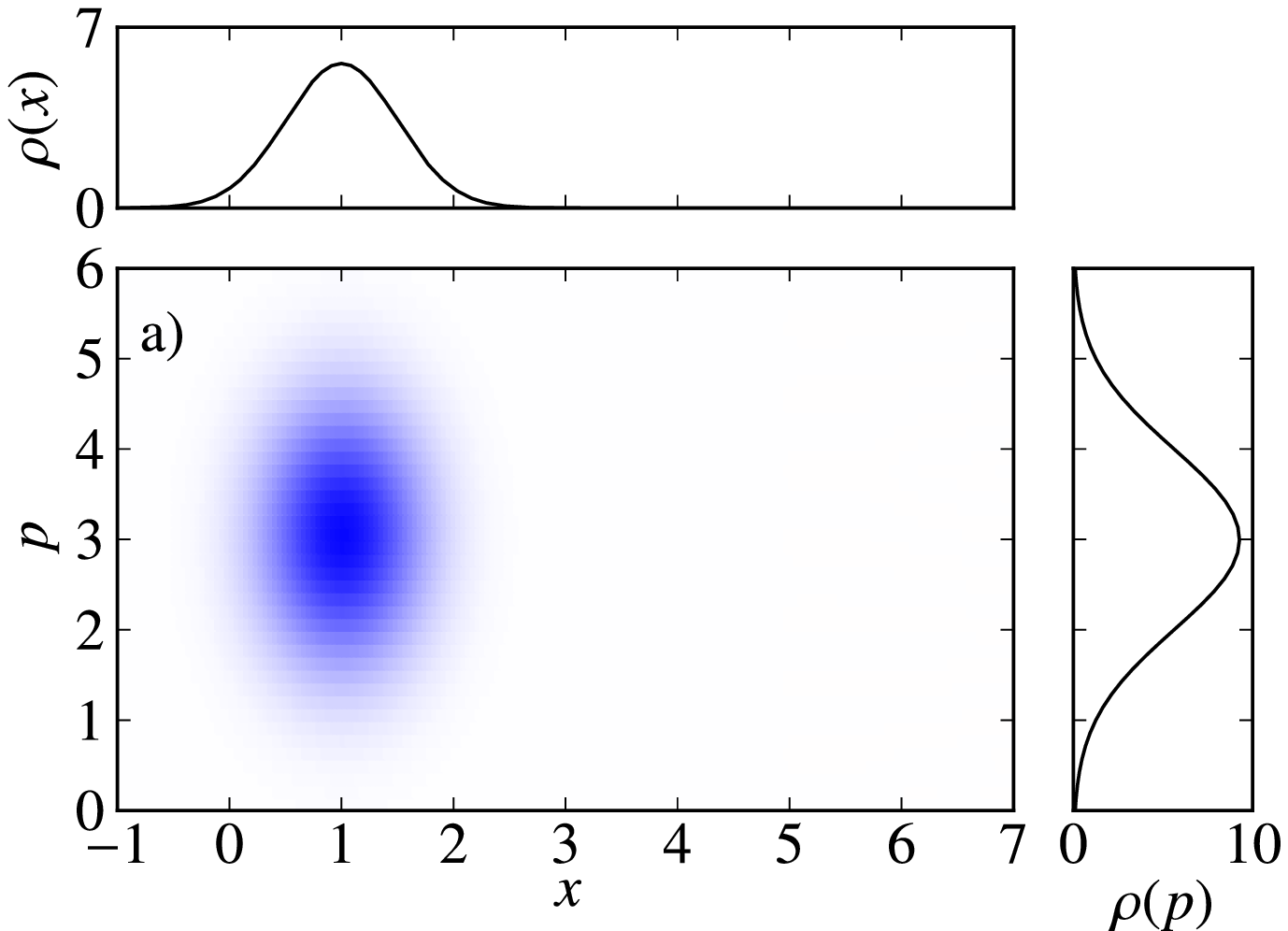}\qquad
  \includegraphics[scale=0.45]{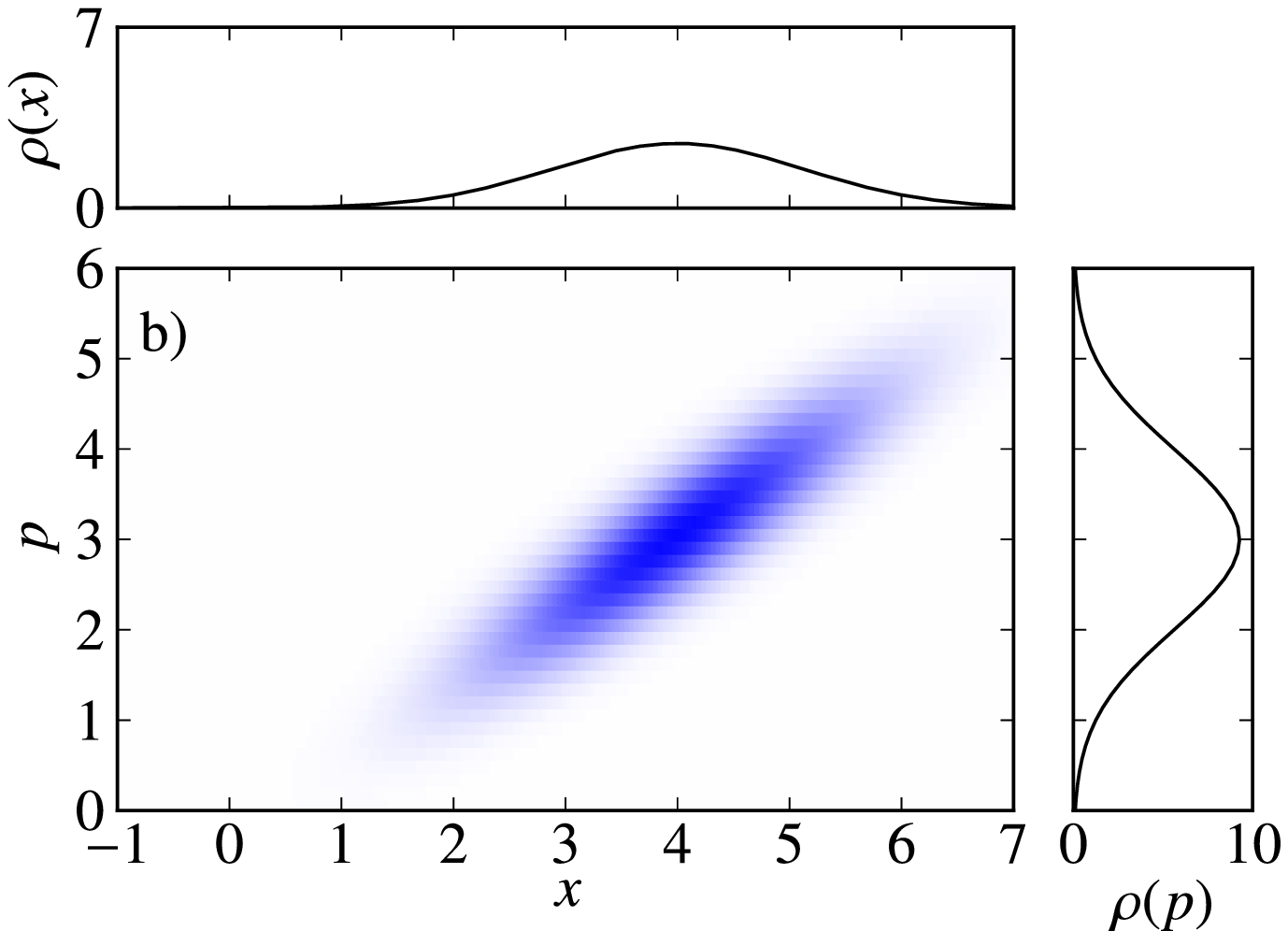}
  \caption{(Color online) Visualization of the Wigner function
    (\ref{eq:w_gauss}) of a Gaussian wave packet by false color plots at
    time $t=0$ (left part) and $t=1$ (right part) with $\bar x=1$, $\bar
    p=3$, and $\sigma=1/2$ (dimensionless units with $\hbar=1$ and
    $m=1$). The panels above the false color plots show the probability
    distribution in position space (\ref{ew:w_marginal_x}) while panels
    right next to the false color plots show the probability
    distribution in momentum space (\ref{ew:w_marginal_p}). Values of
    the Wigner function are indicated by the color-bar at the left.}
  \label{fig:free_particle_w}
\end{figure*}

\section{Visualizing quantum dynamics in phase~space}
\label{sec:vqps} 

After our short review of the Wigner-function's properties in the
previous section, we are now prepared to visualize quantum mechanics in
phase space. This may be accomplished by picturing the Wigner function
itself or quantities that are derived from the Wigner function, e.\,g.,
the Wigner function flow (\ref{eq:Wigner_flow}).

In contrast to the wave function, the Wigner function is a real-valued
quantity. It assigns just a single real value to each point in phase
space, which simplifies the visualization of Wigner functions to some
degree as compared to the visualization of wave functions, which assign
two real values to each point in configuration space.  The dimension of
the phase space---which is twice the configuration-space's
dimension---turns out to be a limiting factor when visualizing Wigner
functions. Therefore, we restrict ourselves in this contribution to
systems with a two-dimensional phase space.  Displaying two-dimensional
real-valued functions is a standard task of scientific visualization and
may be accomplished by various methods, e.\,g., by two-dimensional false
color plots (sometimes called heat maps), by contour line maps, which
show lines of constant values of the Wigner function, or by (pseudo)
three-dimensional surface plots.

Systems with two or more spatial dimensions have a phase space with at
least four dimensions which prohibits visualizing the full Wigner
function. In this case one may in order to reduce the dimensionality,
for example, picture lower dimensional cuts of the Wigner function
through the phase space or projections of the Wigner function onto lower
dimensional surfaces.

\enlargethispage{5ex}%

\subsection{A free wave packet}

The Wigner function of the Gaussian wave packet (\ref{eq:gauss}) is given
by the strictly positive bivariate normal distribution
\begin{equation}
  \label{eq:w_gauss}
  w(x, p, t) = \frac{1}{\pi\hbar}\exp\left(
    -\frac{(x-pt/m-\bar x)^2}{2\sigma^2}
    -\frac{2\sigma^2(p-\bar p)^2}{\hbar}
  \right) \,,
\end{equation}
which is illustrated in Fig.~\ref{fig:free_particle_w} by two false
color plots for two different points in time, for $t=0$ in part~a and
for $t=1$ in part~b, respectively. At time $t=0$, the Wigner function of
the Gaussian wave packet is symmetric in phase space with symmetry axes
parallel to the $x$-axis and the $p$-axis. During the evolution of the
wave packet, the Wigner function gets rotated and stretched along the
$x$-axis in phase space. This corresponds to the familiar wave packet
broadening in position space, see also the probability distribution in
position space (\ref{ew:w_marginal_x}) as shown above the false color
plots in Fig.~\ref{fig:free_particle_w}.

As we pointed out in section~\ref{sec:cps}, the quantum Liouville
equation (\ref{eq:qLiouville}) of a free quantum particle's Wigner
function reduced to the classical Liouville equation
(\ref{eq:Liouville}) for an ensemble of free non-interaction particles
with the Hamiltonian $H(x, p, t)=p^2/(2m)$. Thus, the Wigner function of
a free particle satisfies the relation
\begin{equation}
  \label{eq:w_shift}
  w(x, p, t) = w(x-pt/m, p, 0) \,.
\end{equation}
As a consequence of (\ref{eq:w_shift}), we find the equation
\begin{equation}
  \rho(x, t) = \int w(x, p, t)\,\D p = \int w(x-pt/m, p, 0)\,\D p\,,
\end{equation}
which states that the probability distribution in position space
$\rho(x, t)$ at position $x$ and time $t$ is given by an integral over
the Wigner function at time zero along phase space strips which are
rotated against the $p$-axis by an angle of $-\arctan t/m$. Thus,
measuring the distribution $\rho(x, t)$ of a function of the position
$x$ and the time $t$ allows to reconstruct the full Wigner function
$w(x, p, 0)$ at time zero by means of an inverse Radon transformation
\cite{Leibfried:etal:1998:Shadows_Mirrors,Schleich:2001:Quantum_optics}.
This has also been demonstrated experimentally
\cite{Smithey:etal:1993:Measurement_Wigner_distribution,Kurtsiefer:etal:1997:Measurement_Wigner,Breitenbach:etal:1997:Measurement_quantum_states,Lvovsky:etal:2001:Quantum_State_Reconstruction,Deleglise:etal:2008:Reconstruction}.

\subsection{Characterizing eigen-states}

\begin{figure}[b]
  \centering
  \includegraphics[scale=0.45]{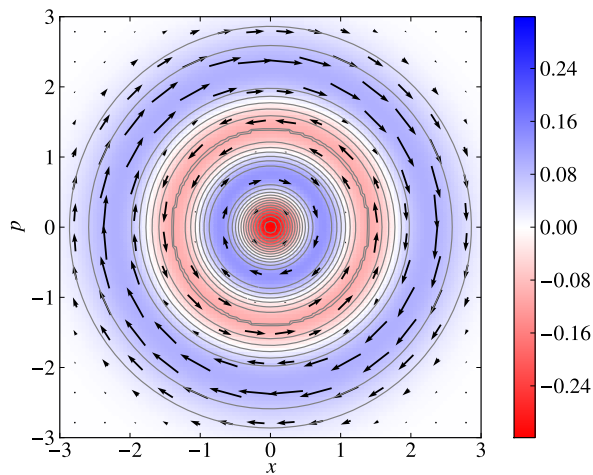}
  \caption{(Color online) False color plot of the Wigner function of the
    harmonic oscillator eigen-function with energy $E=3+1/2$
    (dimensionless units with $\hbar=1$, $m=1$, and $\omega=1$). Gray
    solid lines indicate levels of constant Wigner function values while
    black arrows indicate the steady Wigner function flow
    (\ref{eq:Wigner_flow}). The arrows' length is proportional to the
    flow's magnitude.}
  \label{fig:wigner_ho}
\end{figure}
\begin{figure*}
  \centering
  \includegraphics[scale=0.45]{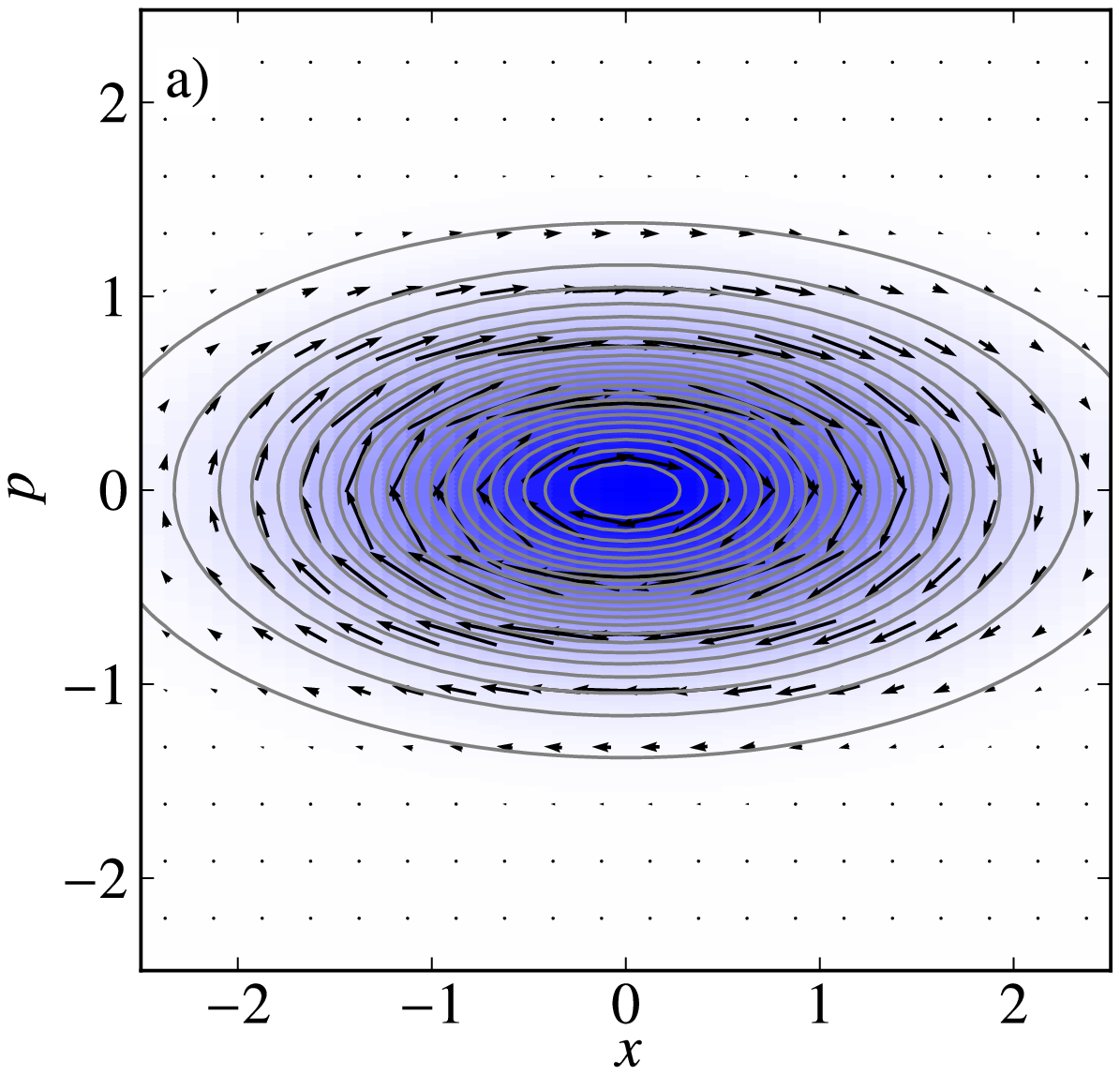}
  \includegraphics[scale=0.45]{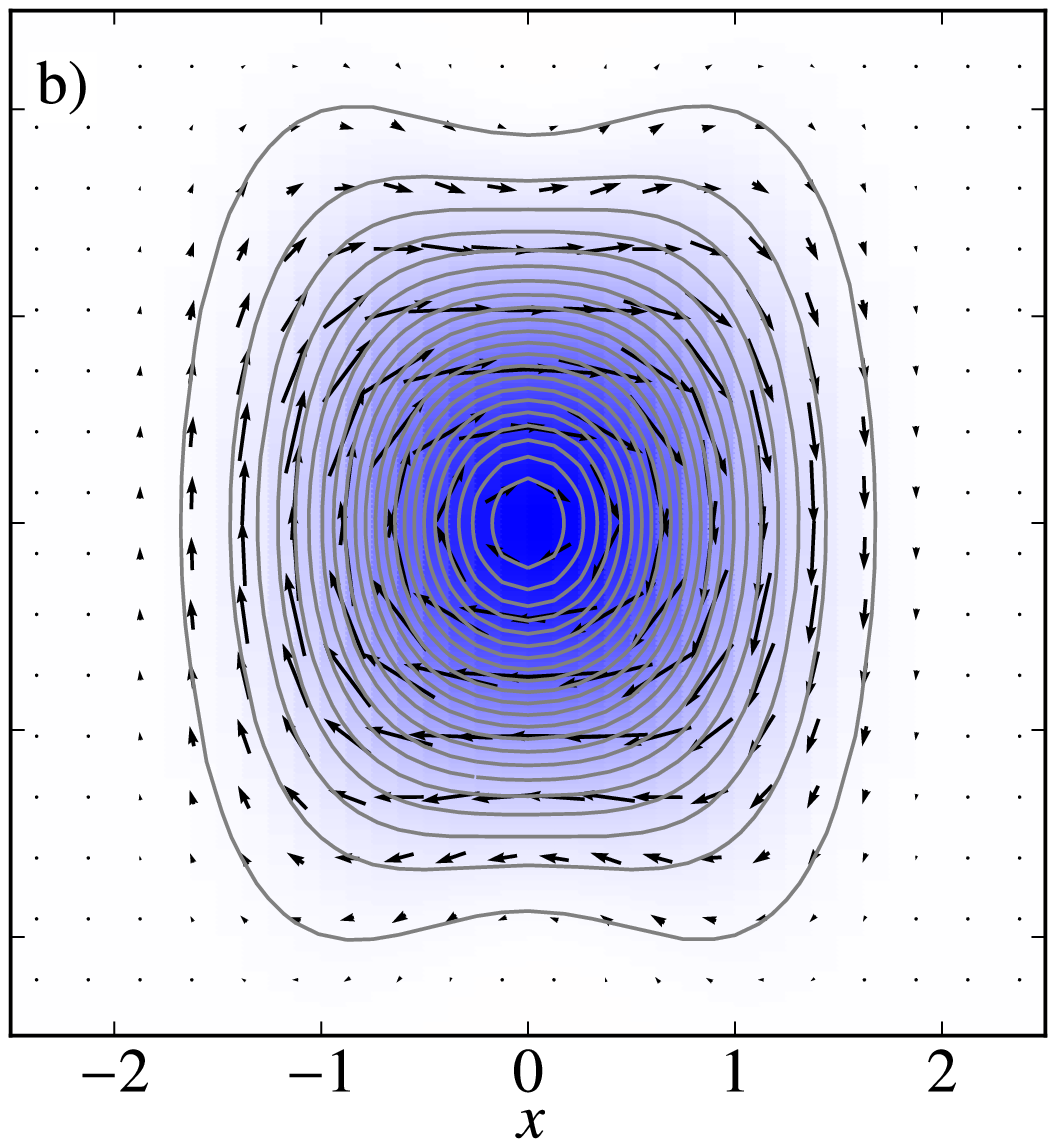}
  \includegraphics[scale=0.45]{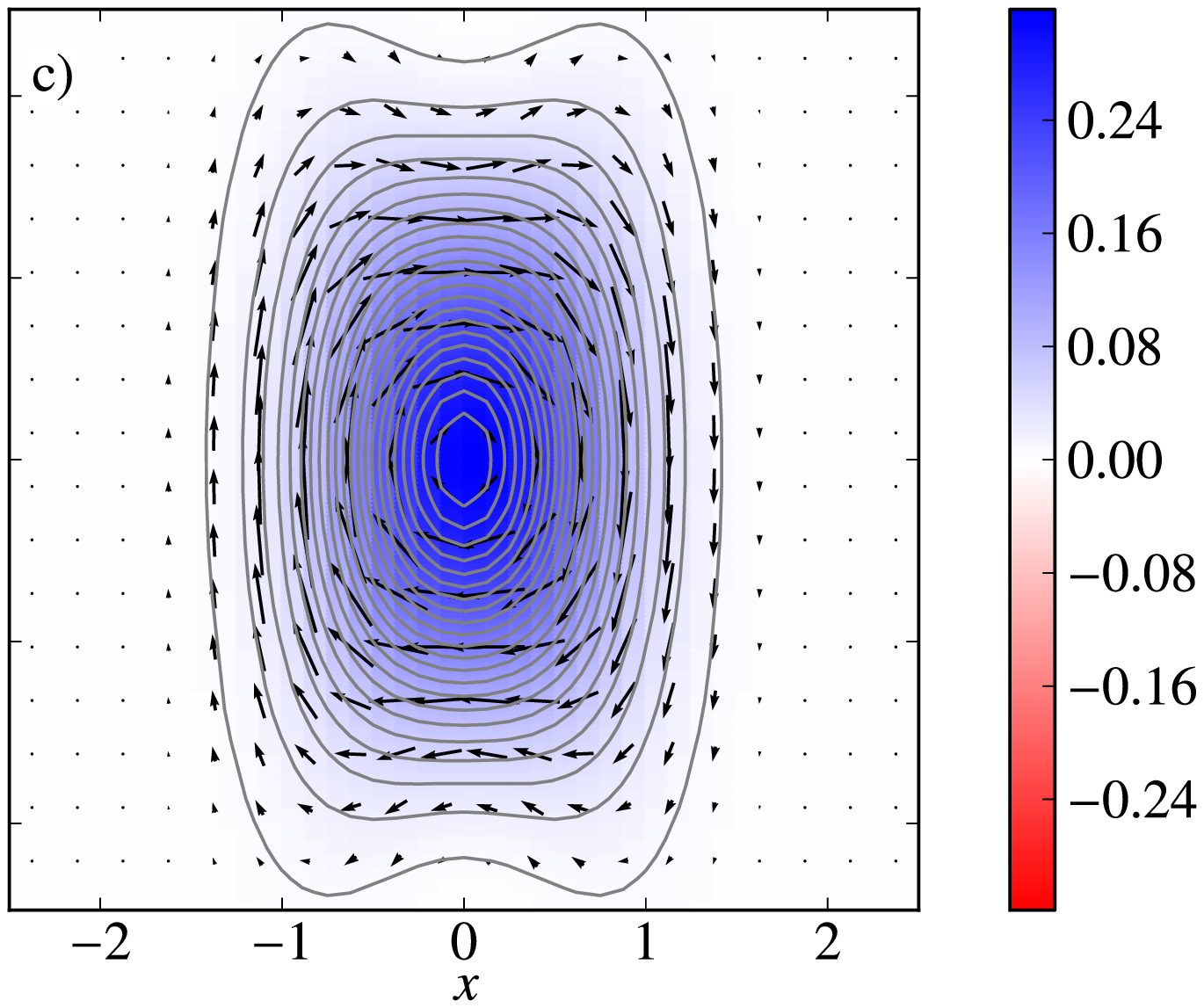}
  \caption{(Color online) Wigner function ground state of the anharmonic
    oscillator potential $V(x)=m\omega^2x^2/2 + \alpha x^4$ for various
    values of the anharmonicity parameter $\alpha$ with $\alpha=0$
    (part~a), $\alpha=1/4$ (part~b), $\alpha=3/4$ (part~c)
    (dimensionless units with $\hbar=1$, $m=1$, and $\omega=1/2$).}
  \label{fig:wigner_aho}
\end{figure*}

The Wigner function eigen-state of some time independent potential
$V(x)$ is determined by the two equations (\ref{eq:Wigner_eigen1}) and
(\ref{eq:Wigner_eigen2}). For the harmonic oscillator potential
$V(x)=m\omega^2x^2/2$ we find
\cite{Hillery:etal:1984:Distribution_functions,Schleich:2001:Quantum_optics}
the eigen-states
\begin{subequations}
  \begin{multline}
    \label{eq:w_ho}
    w_n(x, p) = \\
    \frac{(-1)^n}{\pi\hbar}
    \exp\left(
      - \frac{p^2}{(\hbar\kappa)^2} - (x\kappa)^2
    \right) 
    L_n\left(
      2\frac{p^2}{(\hbar\kappa)^2} + 2(x\kappa)^2
    \right)\,,
  \end{multline}
  and eigen-energies
  \begin{equation}
    E_n = \hbar\omega\left(n+\frac{1}{2}\right)
  \end{equation}
\end{subequations}
with $\kappa=\sqrt{m \omega/ \hbar}$, $n=0,1,\dots$ and $L_n(x)$
denoting the $n$th Laguerre polynomial.

In Fig.~\ref{fig:wigner_ho} we depict the Wigner function of the
harmonic oscillator eigen-state with energy $E=\hbar\omega(3+1/2)$
together with its Wigner function flow (\ref{eq:Wigner_flow}). Because
the $n$th Laguerre polynomial has $n$ positive roots, the Wigner
function of the harmonic oscillator eigen-state oscillates and changes
\mbox{$n$ times} its sign. Note that despite the fact that the Wigner
function (\ref{eq:w_ho}) is time-independent there is a permanent
circular steady Wigner function flow in phase space as indicated by the
arrows in Fig.~\ref{fig:wigner_ho}.

In section~\ref{sec:qps} we demonstrated that for harmonic potentials
quantum \emph{dynamical} effects are absent. This means, the dynamics of
the Wigner function follows the classical Liouville equation because all
quantum terms in the quantum Liouville equation vanish for harmonic
potentials. Also in equation~(\ref{eq:Wigner_eigen}) quantum terms
vanish, which has consequences for the geometry of the Wigner function
flow of the eigen-states of the harmonic oscillator. Using equations
(\ref{eq:Wigner_flow}) and (\ref{eq:Wigner_eigen1}) one can show that
$\vec j(x, p)\cdot\nabla w(x, p)=0$ and, therefore, the Wigner function
flow is always tangential to the lines of constant Wigner function
values. This is a feature that the harmonic oscillator eigen-states
share with all time-independent classical phase space
distributions. Also for classical distributions the flow
(\ref{eq:class_flow}) is tangential to the lines of constant phase space
distribution.

If the potential has non-quadratic terms then the quantum terms in
(\ref{eq:Wigner_eigen}) will affect the structure of the Wigner
eigen-functions and the Wigner function flow and the lines of constant
Wigner function values are no longer tangential to each other. Thus, the
divergence between the direction of the Wigner function flow and the
tangentials to the lines of constant Wigner function values may be
interpreted as a measure how strong the non-classical terms in the
quantum Liouville equation (\ref{eq:qLiouville}) affect the structure of
the eigen-states. 

In Fig.~\ref{fig:wigner_aho} we show the Wigner function ground states
of the anharmonic oscillator potential $V(x)=m\omega^2x^2/2 + \alpha
x^4$ for various values of the anharmonicity parameter $\alpha$. For
this figure the Schr{\"o}dinger ground state wave functions have been
calculated by imaginary time propagation
\cite{Goldberg:Schwartz:1956:imaginary_time_I,Goldberg:Schwartz:1956:imaginary_time_II}
and the Wigner function was obtained by calculating
(\ref{eq:wigner_def}) numerically afterward. In
Fig.~\ref{fig:wigner_aho} the anharmonicity parameter $\alpha$ grows
from left to right starting with $\alpha=0$ (part~a) and increasing to
$\alpha=1/4$ (part~b) and $\alpha=3/4$ (part~c). One clearly sees in
Fig.~\ref{fig:wigner_aho} how the divergence between the direction of
the Wigner function flow and the tangentials to the lines of constant
Wigner function values grows with increasing $\alpha$.

\subsection{Scattering by a potential barrier}

\begin{figure}
  \centering
  \includegraphics[scale=0.45]{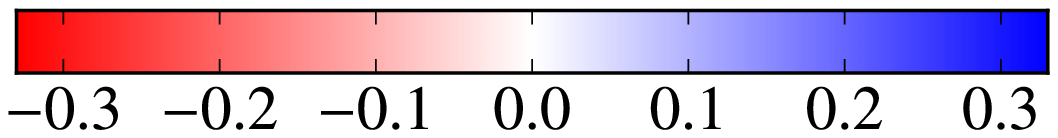}
  \includegraphics[scale=0.45]{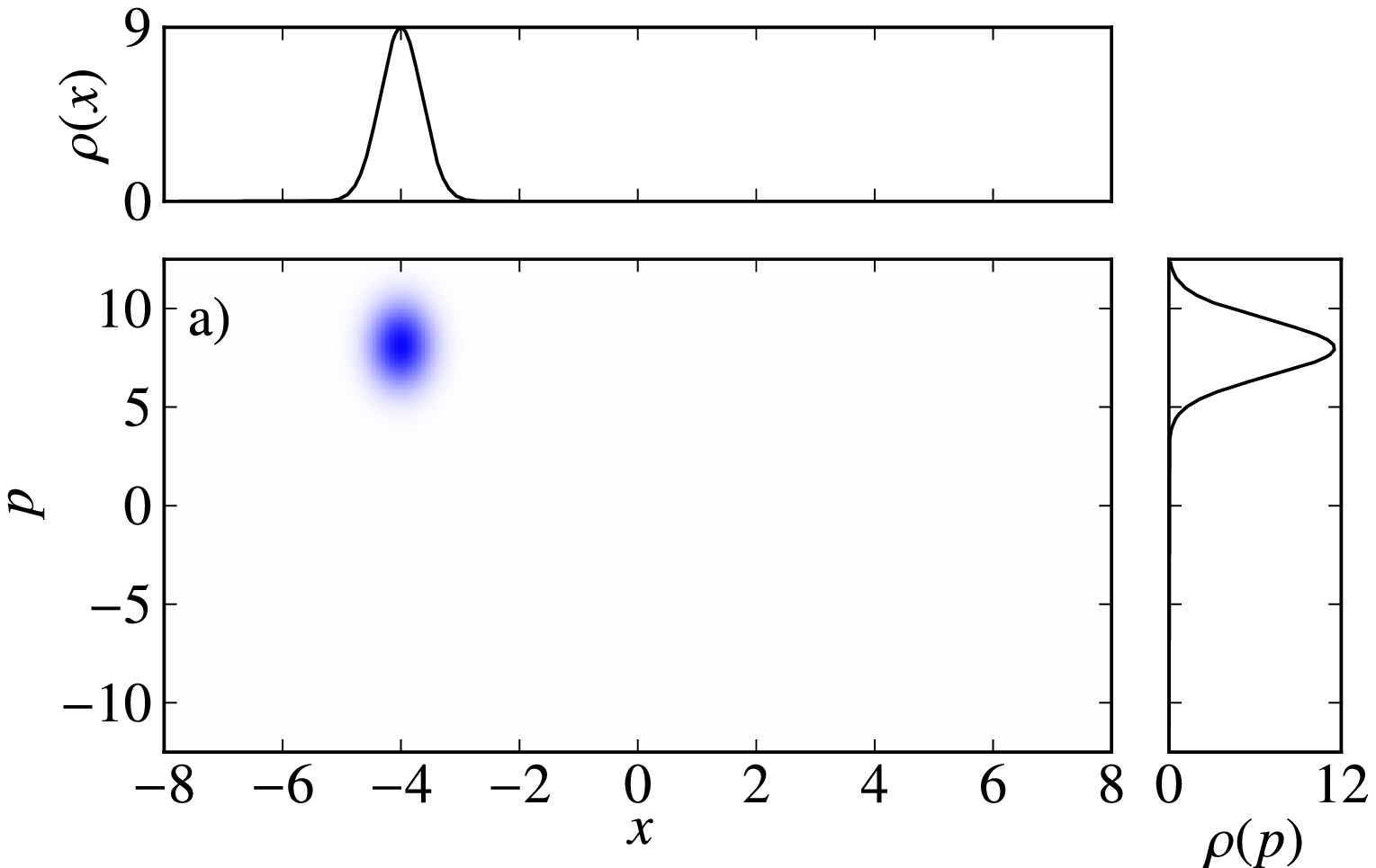}
  \includegraphics[scale=0.45]{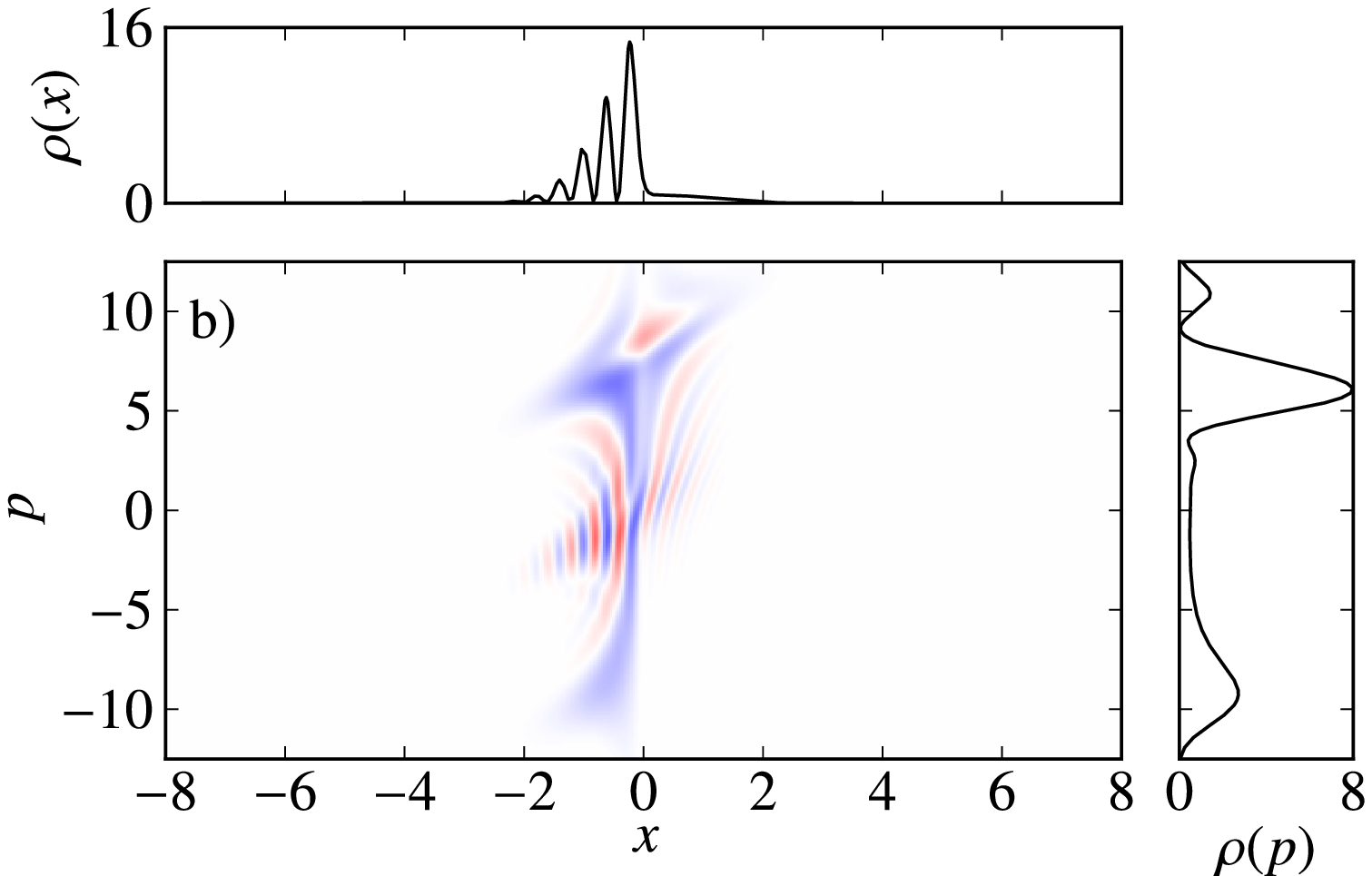}
  \includegraphics[scale=0.45]{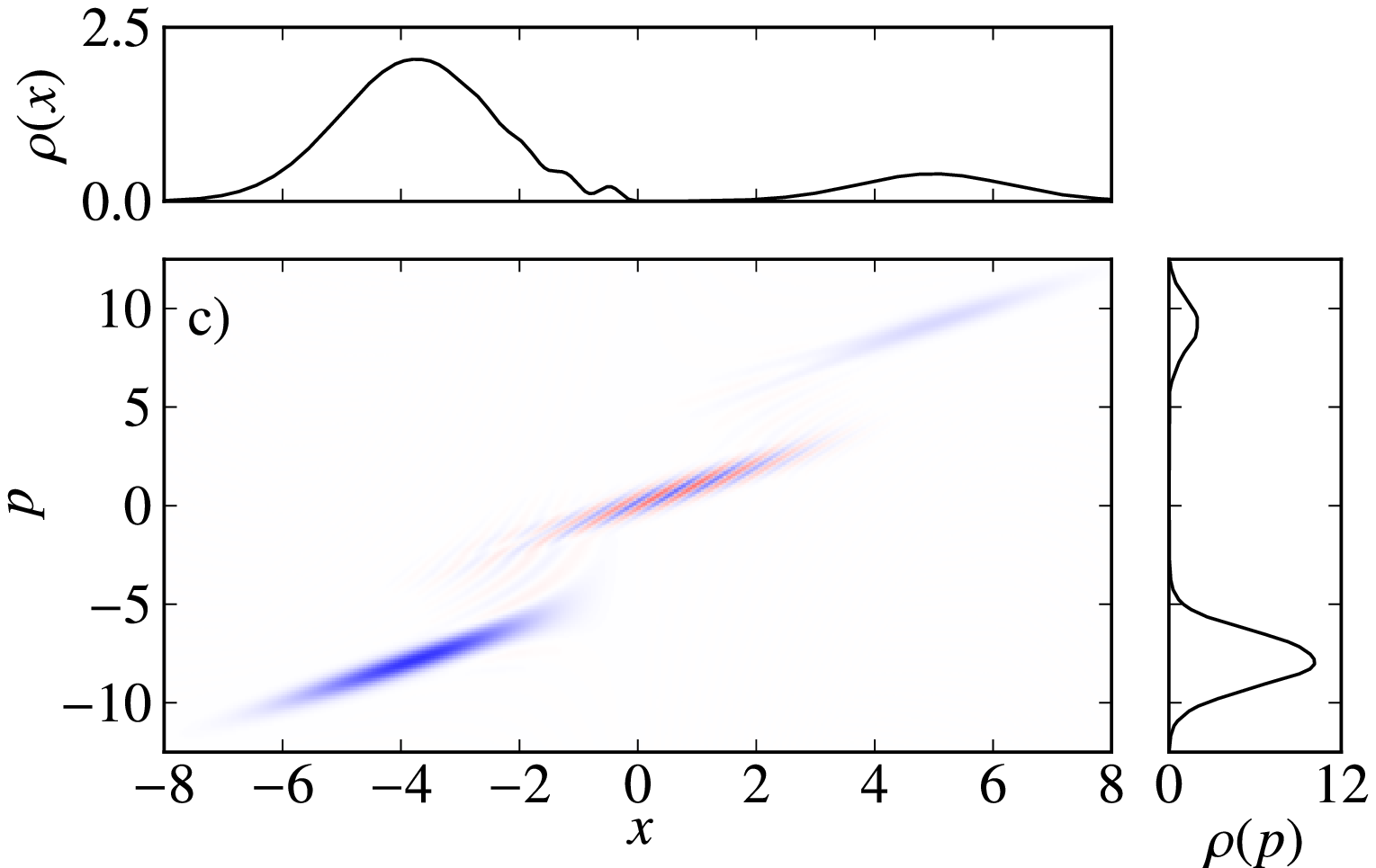}
  \includegraphics[scale=0.45]{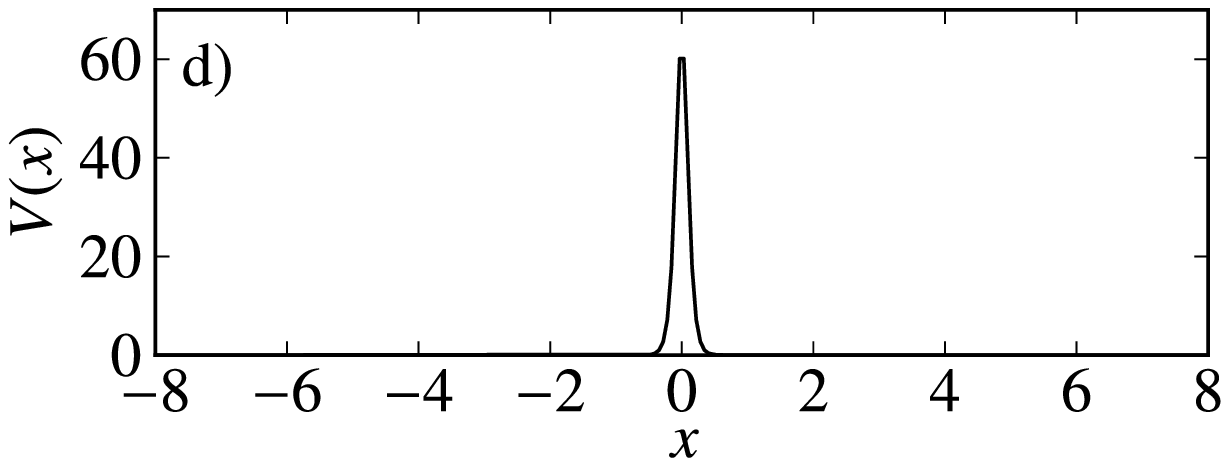}
  \caption{(Color online) Scattering of a wave packed by a potential
    barrier (\ref{eq:barrier}) in phase space. Potential parameters are
    $V_0=64$ and $\Delta =1/8$, the wave-packet's initial mean momentum
    is $8$, all values in dimensionless units with $\hbar=1$ and
    $m=1$. The three upper panels show the Wigner function, the position
    space probability distribution and the momentum space probability
    distribution at different times; $t=0$ (part~a), $t=1/2$ (part~b),
    and $t=1$ (part~c). Part~d illustrates the scattering potential
    (\ref{eq:barrier}).}
  \label{fig:barrier}
\end{figure}

Scattering of a wave packed by a potential barrier is a well known
textbook system to exemplify quantum effects. Depending on the
potential's shape and the initial condition of the wave packet some part
of the wave packet may be reflected while the other part is transmitted;
a process that has no analogon in classical physics of point-like
particles. 

The quantum dynamical scattering dynamics is commonly visualized by the
wave function in position space
\cite{Goldberg:etal:1967:Motion_Pictures,Brandt:Dahmen:2001:Picture_book}
and sometimes in momentum space \cite{Goldberg:etal:1968:Scattering}.
Figure~\ref{fig:barrier} illustrates scattering of an initial Gaussian
wave packet by a potential barrier in phase space at different points in
time. The scattering potential barrier is located around \mbox{$x=0$}
and its shape is given by
\begin{equation}
  \label{eq:barrier}
  V(x)=V_0
  \left(\tanh\left(\frac{x}{\Delta}\right)+1\right)
  \left(\tanh\left(-\frac{x}{\Delta}\right)+1\right) \,,
\end{equation}
see also Fig.~\ref{fig:barrier}~d.  For this figure the Schr{\"o}dinger wave
functions was propagated by a Fourier split operator method
\cite{Feit:etal:1982:spectral_method} and the Wigner function was
obtained by calculating (\ref{eq:wigner_def}) numerically afterward.

The system in Fig.~\ref{fig:barrier} starts at time $t=0$
(Fig.~\ref{fig:barrier}~a) with an asymptotically free Gaussian wave
packet. Here we consider a thin potential with height $V_0=64$ and
characteristic width $\Delta=1/8$. The wave packet's initial kinetic
energy equals $E_\mathrm{kin}=32$. Accordingly to the laws of classical
physics the kinetic energy is too small to cross the barrier.  As the
wave packet moves towards the potential barrier it starts to interact
with the barrier. Due to the non-classical terms of the quantum
Liouville equation the Wigner function starts to develop regions with
negative values such that at about $t=1/2$ significant interference
patterns have emerged, see Fig.~\ref{fig:barrier}~b. At about $t=1$ the
wave packet has split into two parts that are traveling into opposite
directions in position space, a reflected part and a transmitted part
that has tunneled trough the potential barrier. In phase space this
corresponds to two isolated wave packets with approximately Gaussian
shape moving into opposite directions, see also
Fig.~\ref{fig:barrier}~c. However, there is also a region with non-zero
Wigner function between these two packets that exhibits strong
oscillations featuring positive and negative values reviling the quantum
nature of the coherent superposition of the two wave packets
\cite{Kurtsiefer:etal:1997:Measurement_Wigner,Deleglise:etal:2008:Reconstruction}.

\section{Conclusions}

In this contribution we have explored ways to visualize quantum
mechanics in phase space by means of the Wigner function and the Wigner
function flow. The mathematical formalism of quantum mechanics in phase
space in terms of the Wigner function resembles classical Hamiltonian
mechanics of non-interacting classical particles. Thus, teaching quantum
mechanics in phase space may have some advantages when moving from
classical to quantum physics. Non-classical features are highlighted in
phase space, for example by negative values of the phase space
distribution or the divergence between the direction of the Wigner
function flow and tangentials to the lines of constant Wigner function
values of eigen-states.

\begin{acknowledgments}
  N.\ R.\ I.\ would like to express her appreciation to the organizers of
  the XVth International Summer Science School of the city of Heidelberg
  and to the Max-Planck-Institut f{\"u}r Kernphysik for its hospitality.
\end{acknowledgments}

\bibliography{phase_space_vis}

\end{document}